\documentclass[review,12pt]{elsarticle}

\usepackage{extsizes}
\usepackage[version=3]{mhchem}
\usepackage[left=2.5cm, right=2.5cm, top=1.785cm, bottom=2.0cm]{geometry}
\usepackage{times,mathptmx}
\usepackage{gensymb}
\usepackage{sectsty}
\usepackage{epsfig}
\usepackage{lastpage}
\usepackage{float}
\usepackage{fnpos}
\usepackage[english]{babel}
\usepackage{array}
\usepackage{droidsans}
\usepackage{charter}
\usepackage[T1]{fontenc}
\usepackage[usenames,dvipsnames]{xcolor}
\usepackage{setspace}
\usepackage[compact]{titlesec}
\usepackage{array}
\usepackage{color}
\usepackage{textcomp}
\usepackage{amsmath}
\usepackage{amssymb}
\usepackage{amsthm}

\usepackage{epstopdf}

\makeatletter
\def\ps@pprintTitle{%
     \let\@oddhead\@empty
     \let\@evenhead\@empty
     \def\@oddfoot{}%
		 \let\@evenfoot\oddfoot}
\makeatother

\begin{document}

\begin{frontmatter}

\title{Relaxation dynamics, Softness and Fragility of Microgels with Interpenetrated Polymer Networks}

\author[ISC,DIP]{Valentina Nigro\footnote{Corresponding author:valentina.nigro@uniroma1.it}}
\author[ISC,DIP]{Barbara Ruzicka\footnote{Corresponding author:barbara.ruzicka@roma1.infn.it}}
\author[CNRS,ESRF]{Beatrice Ruta}
\author[ESRF]{Federico Zontone}
\author[ISOF]{Monica Bertoldo}
\author[ISC]{Elena Buratti}
\author[ISC,DIP]{Roberta Angelini\footnote{Corresponding author:roberta.angelini@roma1.infn.it}}

\address[ISC]{Istituto dei Sistemi Complessi del Consiglio Nazionale delle Ricerche (ISC-CNR), sede Sapienza, Pz.le Aldo Moro 5, I-00185 Roma, Italy}
\address[DIP]{Dipartimento di Fisica, Sapienza Universit$\grave{a}$ di Roma, I-00185, Italy}
\address[CNRS]{France Univ Lyon, Universit$\acute{e}$ Claude Bernard Lyon 1, CNRS, Institut Lumi$\grave{e}$re Mati$\grave{e}$re, Villeurbanne, France}
\address[ESRF]{ESRF The European Synchrotron, CS40220, 38043 Grenoble Cedex 9, France}
\address[ISOF]{Istituto per la Sintesi Organica e la Fotoreattivit$\grave{a}$ del Consiglio Nazionale delle Ricerche (ISOF-CNR), via P. Gobetti 101, 40129 Bologna, Italy}

\begin{abstract}
Microgels are elastic and deformable particles with a hybrid nature between that of polymers and colloids and unconventional behaviours with respect to hard colloids.
We investigated the dynamics of a soft microgel made of interpenetrated polymer networks of PNIPAM and PAAc by means of coherent X-ray and light scattering techniques. By varying the particle softness through PAAc content we can tune at wish the fragility of IPN microgels.
Interestingly we find the occurrence of a dynamical crossover at a critical weight concentration which leads to an evolution of the structural relaxation time from a super-Arrhenius to a slower than Arrhenius behaviour, a minimum for the shape parameter of intensity autocorrelation function and the emerging of distinct anomalous mechanisms for particle motion.
This complex phenomenology can be described by a Fickian diffusion at very low concentrations, an effective non Fickian anomalous diffusion at intermediate values and a ballistic motion well described within the Mode Coupling Theory.
\\
\end{abstract}

\end{frontmatter}

\section{Introduction}

In the last decades many works have been focused on the study of the glass transition in complex systems aiming at understanding differences and similarities between structural glasses (SG) and colloidal glasses (CG)~\cite{SciAdvPhys2005,TrappeCOCIS2004,PoonCOCIS1998,ZaccarelliJPCM2007}. 
In the case of structural glasses, as the liquid is rapidly quenched below its melting temperature, it undergoes a glass transition avoiding nucleation and crystallization. In colloids the transition is usually triggered by volume fraction $\phi$ or waiting time $t_w$ whose increase drives the systems in an out-of-equilibrium configuration.
Changing control parameters like temperature, packing fraction or aging time, the dynamics of both structural and colloidal glasses slows down enormously and is accompanied by a dramatic increase of the characteristic relaxation time ($\tau(T)$, $\tau(\phi)$, $\tau(t_w)$) by several orders of magnitude up to the glassy state.
The most direct way to access microscopic information on the relaxation dynamics is to look at the evolution of the intensity autocorrelation functions which are well described by the Kohlrausch-Williams-Watts (KWW) expression 
$F(Q,t) \propto exp(-(t/\tau(Q))^{\beta})$
 where $\tau$ is an 
"effective" relaxation time and $\beta$ measures the
distribution of relaxation times and assumes generally positive values $\beta$<1 (stretched behaviour). Recently an anomalous dynamics, characterized by a shape parameter $\beta$>1 (compressed behaviour), has emerged in out of equilibrium states of different materials like colloidal glasses~\cite{BellourPRE2003, BandyopadhyayPRL2004, SchosselerPRE2006, AngeliniSM2013, AngeliniNC2014, KwasniewskiSM2014, AngeliniCSA2014, AngeliniCSA2015, PastoreSciRep2017}, gels~\cite{CipellettiPRL2000, ChungPRL2006, GuoJCP2011,  OrsiPRL2012, CristofoliniCOCIS2014, ManselSM2015}, supercooled liquids~\cite{CaronnaPRL2008, ConradPRE2015}, metallic glasses~\cite{RutaPRL2012, EvensonPRL2015}, polymeric systems~\cite{FalusPRL2006, NarayananPRL2006, SrivastavaPRE2009}, ceramics~\cite{BalitskaCP2018} and investigated also in theoretical~\cite{BouchaudEPJE2001, Bouchaud2008, FerreroPRL2014, NicolasArXiv2018} and simulation~\cite{MorishitaJCP2012, BouzidNatCom2017, ChaudhuriPRE2017, PastoreCSA2017, WeiNatCom2018} works.  The existence of stretched and/or compressed correlation functions is currently a very debated issue also faced in this work.
Despite of many studies on the glass transition in SG and CG several questions are still open: Is the \textit{structural relaxation} behaviour peculiar of the \textit{specific investigated} system? How does it depend on the \textit{control parameters}? What is the \textit{$\beta$ parameter} behaviour? What happens to the diffusive \textit{dynamics} of the equilibrium liquid? Here, combining X-ray Photon Correlation Spectroscopy (XPCS) and Dynamic Light Scattering (DLS), we aim to contribute to these fundamental questions investigating the behaviour of the structural relaxation time and $\beta$ exponent for a soft colloid. Recent theoretical works on these topics \cite{ScheerACSNano2017, GnanNat2019} and experimental studies on IPN \cite{MattssonNature2009} and PNIPAM \cite{PhilippePRE2018, MohantySciRep2017} microgels addressed some of these points.
Soft colloids are an interesting class of glass-formers that, at variance with hard sphere-like colloids, provide a good tunability of particle softness giving rise to unconventional phase-behaviours~\cite{LikosPhysRep2001, RamirezJPCM2009, VlassopoulosCOCIS2014}.
They are indeed aqueous suspensions of nanometre- or micrometre-sized hydrogel particles sensitive to external stimuli whose dimension and effective volume fraction can be varied by changing external parameters such as temperature and/or pH. 

The most studied responsive microgels are those based on the thermo-sensitive poly(N-isopropylacrylamide) (PNIPAM), with a reversible Volume Phase Transition (VPT) around 305 K. These systems have been largely investigated both theoretically and experimentally~\cite{PeltonAdvColloid2000, FernandezBook2011, KargLangmuir2019, WuPRL2003, LyonRevPC2012, PaloliSM2013, SeekellSM2015, GnanMacromol2017, PhilippePRE2018, BergmanNatComm2018, TavagnaccoJPCL2019, NinarelloMacromol2019, ScottiMacromol2019}. In this work we study a PNIPAM-based microgel with a second interpenetrated polymer network (IPN) of poly(acrylic acid) (PAAc)~\cite{XiaLangmuir2004, ZhouBio2008, MattssonNature2009, MaColloidInt2010, XingCollPolym2010, LiuPolymers2012, LiMaterEng2013, NigroJNCS2015, NigroSM2017, NigroJML2019} that provides additional pH sensitivity, topological constraints to the particles and extra charges to the system. In this way the IPN microgel softness can be controlled by synthesis varying the amount of poly(acrylic acid). 

\section{Experimental Methods}
\label{Experimental Methods}

The dynamic structure factor has been investigated  
at different scattering vectors $Q$, weight concentrations $C_w$ and PAAc content $C_{PAAc}$ through combined XPCS and DLS techniques.

XPCS measurements were performed at ID10 beamline of ESRF in Grenoble using a
partially coherent X-ray beam with a photon energy of 21 keV. A
series of scattering images were recorded by CdTe Maxipix detector (photon counting). The ensemble averaged intensity autocorrelation
function $g_2(Q, t)=\frac{\langle\langle
I(Q,t_0)I(Q,t_0+t)\rangle_p \rangle}{\langle \langle
I(Q,t_0)\rangle_p \rangle}$, where $\langle...\rangle_p$ is the
ensemble average over the detector pixels mapping onto a single
$Q$ value and $\langle...\rangle$ is the temporal average over
$t_0$, was calculated by using a standard multiple $\tau$
algorithm~\cite{MadsenNJP2010, LehenyCOCIS2012}. XPCS data were complemented by
DLS measurements performed at the CNR-ISC laboratory. The monochromatic and polarized beam
emitted from a solid state laser (100 mW at $\lambda$ = 642 nm) was
focused on the sample placed in a cylindrical VAT for index
matching and temperature control. 
The scattered intensity was simultaneously collected by single mode optical fibers at five different scattering
angles, namely $\theta$=30\textdegree, 50\textdegree, 70\textdegree,
90\textdegree, 110\textdegree, corresponding to different
scattering vectors $Q$, according to the relation $Q$=(4$\pi$n/$\lambda$) sin($\theta$/2).

The investigated samples were IPN microgels synthesized by a sequential free radical polymerization method \cite{XiaLangmuir2004} starting from a PNIPAM dispersion prepared as described in Ref.\cite{NigroJCP2015, NigroCSA2017}. AAc and N,N'-methylene-bis-acrylamide (BIS) were added into the preformed PNIPAM microgels at $T$=294 K, allowing the growth of the PAAc network inside the particles. The mixture was diluted with ultrapure water and transferred into a four-necked jacketed reactor kept at 294 K and deoxygenated by bubbling nitrogen for 1 h. N,N,N',N'-tetrame thylethylenediamine (TEMED) was added and polymerization started with ammonium persulfate. Five samples at different PAAc/PNIPAM ratio composition were prepared by stopping the reaction at the suitable degree of conversion of AAc. The samples were purified by dialysis against distilled water with frequent water changes for 2 weeks. The synthesized particles were analysed by ATR FT-IR and $^1$H-NMR spectroscopies, as well as by elemental analysis to assess their chemical composition and the exact PAAc content \cite{VillariCPC2018}. The obtained samples have the following PAAc weight concentrations: $C_{PAAc}$ = 2.6 \%; $C_{PAAc}$ = 10.6 \%; $C_{PAAc}$ = 15.7 \%; $C_{PAAc}$ = 19.2 \%; $C_{PAAc}$ = 24.6 \%.

Samples at different concentrations were obtained by dilution at pH close to 5.5.    
Measurements were performed on aqueous suspensions of IPN microgels at the five PAAc content indicated above, fixed temperature above the VPTT ($T$=311 K), different weight concentrations $C_w$=(0.05 $\div$ 5) \% and acidic pH (pH $\sim$ 5.5) in the $Q$ range $Q$=(0.006 $\div$ 0.063) nm$^{-1}$ below the peak of the static structure factor. 
We did not observe any indication of aging during the measurements.
The particles radii of the investigated samples at $T$=311 K are: $C_{PAAc}$=2.6 \% R=(26 $\pm$ 1) nm, $C_{PAAc}$=10.6 \% R=(52 $\pm$ 2) nm, $C_{PAAc}$=15.7 \% R=(68 $\pm$ 3) nm, $C_{PAAc}$=19.2 \% R=(89 $\pm$ 2) nm, $C_{PAAc}$=24.6 \% R=(130 $\pm$ 2) nm; they increase with PAAc content as reported in Ref.~\cite{NigroJCIS2019}. An example of the temperature behaviour of hydrodynamic radius for $C_{PAAc}$=24.6 \% and $C_w$=0.01 \% is reported in Fig.~\ref{fig:g2Z}(b). 

\section{Results and Discussion}

\begin{figure*}[!t]
\centering
\includegraphics[width=10cm]{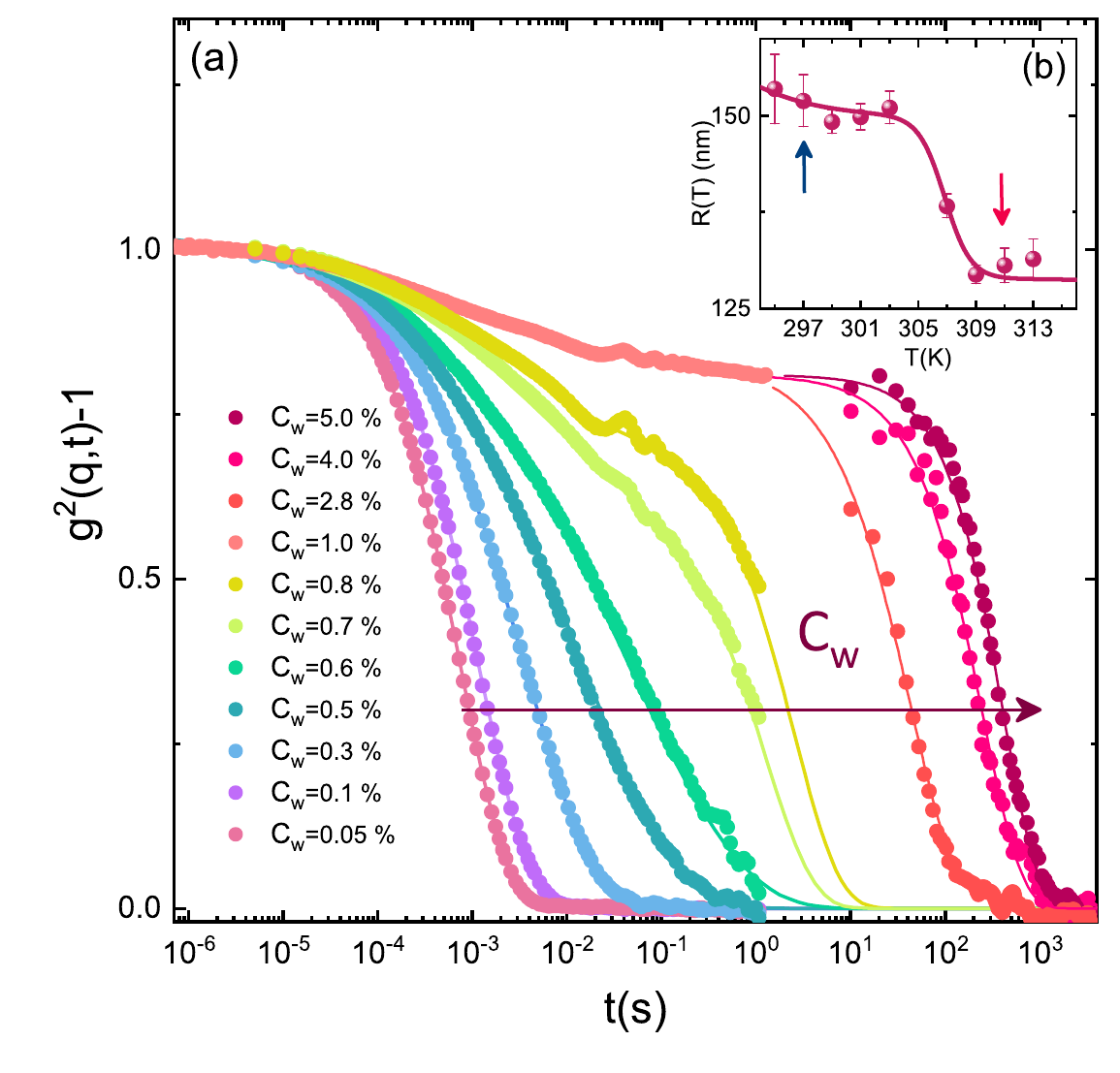}
\caption{(a) Normalized intensity autocorrelation functions of IPN microgel particles at $C_{PAAc}$=24.6 \%, $T$=311 K and $Q$=0.022 nm$^{-1}$ as a function of concentration and (b) Temperature behaviour of the hydrodynamic radius at $C_w$=0.01 \%, blue and red arrows indicate $T$=297 K and $T$=311 K respectively. }
\label{fig:g2Z}
\end{figure*}

Figure~\ref{fig:g2Z} shows DLS and XPCS intensity
autocorrelation functions at 0.05 \% $\leq$ C$_w$ $\leq$ 1.0 \% and  1.0 \% < C$_w$ $\leq$ 5.0 \% respectively, for IPN microgels with fixed PAAc content ($C_{PAAc}$=24.6 \%) at different concentrations $C_w$ and fixed temperature $T$=311 K (red arrow in Fig. \ref{fig:g2Z}(b)) when the particles are in the shrunken state. 
At low $C_w$ the intensity scattering functions are well described by a monomodal decay capturing the particle diffusion in the high dilution limit. As the particle concentration increases a dynamical decoupling is observed and a two-step relaxation, characteristic of glass-forming systems approaching the glass transition, comes out. The monomodal decay for $C_w$ < 0.6 \% and the final decay for $C_w$ $\geq$ 0.6 \%  are well described by the Kohlrausch-Williams-Watts expression~\cite{KohlrauschAnnPhys1854, WilliamsFaradayTrans1970}:

\begin{equation}\label{eq1}
g_2(Q,t)= b[(Aexp(-(t/\tau(Q))^{\beta}))^2+1]
\end{equation}

where $ b \cdot A^2 $ is the coherence factor, $\tau$ the structural relaxation time and $\beta$ the shape parameter, representing a phenomenological hallmark of glass forming liquid dynamics. 
The fits are shown as full lines in Fig.~\ref{fig:g2Z}. 

\begin{figure*}[!t]
\centering
\includegraphics[trim={0cm 2cm 2cm 0cm},clip,width=17cm]{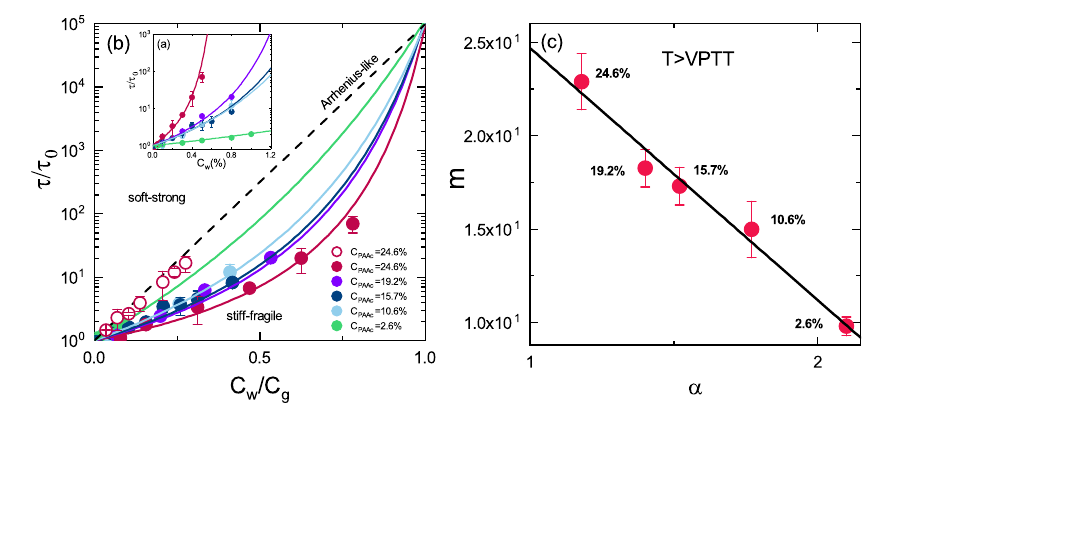}
\caption{\label{fig:tauDLS} (a) Normalised structural relaxation time as a function of weight concentration at the indicated PAAc contents and $T$=311 K. Solid lines are fits through Eq.(\ref{VFT}). (b) Angell plot for the normalized structural relaxation time 
versus normalized weight concentration at $T$=311 K (closed symbols) compared with data at $T$=297 K for a sample at 24.6 \% of PAAc as an example (open symbols). 
Solid lines are fits according to Eq.(\ref{VFT}). (c) Fragility index m 
as a function of swelling ratio related to particles softness.}
\end{figure*}

The normalised structural relaxation time is reported in Fig.\ref{fig:tauDLS}(a) for different $C_{PAAc}$ as a function of C$_w$: with increasing PAAc content a steeper growth and a divergence occurring at lower concentrations are found.
This can be related to the increased interactions due to COOH groups belonging to PAAc chains as deeply discussed in Ref.~\cite{NigroJCIS2019}. Our data are well described by the super-Arrhenius expression:

\begin{equation} 
\tau=\tau_0 \: exp(\; \frac{D_{C_w}C_w}{C_{w0}-C_w}) \label{VFT}
\end{equation}

where $C_{w0}$ sets the apparent divergence, $D_{C_w}$ controls the growth of the structural relaxation time and $\tau_0$ is the characteristic structural relaxation time in the high dilution limit. 

In Fig.\ref{fig:tauDLS}(b) the data for each PAAc content are rescaled by content are rescaled by $C_g$, which is the $C_w$ at which the relaxation time is 100 s. One can observe that at T=311 K the curves show a more and more fragile behaviour by increasing PAAc content while a strong (Arrhenius-like) behaviour is recovered only in the case of microgels at T=297 K (blue arrow in Fig.\ref{fig:g2Z}(b)), for all the investigated PAAc contents. As an example only $C_{PAAc}$=24.6 \% (open symbols) is reported in Fig.\ref{fig:g2Z}(b). This result points out that in microgels, by changing PAAc content and/or temperature, different dynamical behaviours can be achieved as in the case of molecular glass formers \cite{AngellPNAS1995}.

Recently, the unifying concept of fragility has been extended to colloidal suspensions~\cite{MattssonNature2009, SeekellSM2015, BeheraPRM2017, ScheerACSNano2017, GnanNat2019} and many efforts have been devoted to understand the effect of softness on fragility \cite{ScheerACSNano2017, GnanNat2019, MattssonNature2009, SenguptaJCP2011, ShiJCP2011}. 
In this context the slope of the data at $C_w=C_g*$ can be used to define a kinetic fragility as:

\begin{equation} 
m= \left[\frac{\partial log \tau}{\partial (C_w/C_g)}\right]_{C_w=C_g} \label{fragility}
\end{equation}
\\
Since the pioneering study of Ref. \cite{MattssonNature2009} the connection between softness and fragility in soft colloids is still largely debated. 
A recent work on PNIPAM microgels (Ref. \cite{PhilippePRE2018}) states that regardless of their softness, soft colloids exhibit the same fragile-like behaviour of hard colloids. In agreement with these findings, we observe a fragile-like behaviour only at high temperature, however we show that fragility can be also finely controlled by tuning the amount of PAAc interpenetrated within PNIPAM network and therefore particles softness. On the contrary, at low temperature we find a strong-like behaviour as also reported in Ref. \cite{NigroSM2017}. A simple theoretical model (Ref.\cite{ScheerACSNano2017}) has found that depending on osmotic pressure, presence of ions, dimension of particles, osmotic deswelling may be significant, giving rise to strong rather than fragile glasses. 
To better investigate the dependence of fragility on softness, we define the swelling ratio as:

\begin{equation}
\alpha = \frac{R_H^{swollen}}{R_H^{shrunken}}
\end{equation}

where $R_H^{swollen}$ = R($T$=297 K) and $R_H^{shrunken}$ = R($T$=311 K) (blue and red arrows in Fig.\ref{fig:g2Z}(b) respectively). Particles with higher softness can shrink more, providing a larger value of $\alpha$ that decreases with increasing PAAc content as reported in Ref. \cite{NigroJCIS2019}. These values are strictly related to the Young modulus of particles as shown in Ref. \cite{Angelini2020}. 
Plotting the fragility $m$ as a function of $\alpha$ (Fig.\ref{fig:tauDLS}(c)) we find a linear decrease with increasing particle softness in surprisingly agreement with very recent theoretical results on the role of deformation in soft colloids \cite{GnanNat2019}. A clear picture of the relation between softness and fragility is thus provided: the higher is the PAAc content, the stiffer are the particles, yielding thus to more fragile systems. 
These results compared to previous works \cite{MattssonNature2009, PhilippePRE2018, ScheerACSNano2017} corroborate the idea that soft colloids can give rise to both fragile and strong behaviours. In our system the interpenetration of PAAc determines an increase of topological constraints, an increase of particle dimensions and extra charges. IPN microgels represent therefore a good prototype to investigate the role of different parameters varied ad hoc in a single system. 
To broaden the investigated dynamical range for $C_w$>$C_g$, we focus our attention on particles with the highest PAAc content (C$_{PAAc}$=24.6 \%), as they allow to get closer to the critical weight concentration (see Fig.\ref{fig:tauDLS}(b)).

\begin{figure*}[t!]
\centering
\includegraphics[trim={0cm 2cm 0cm 0cm},clip,width=17cm]{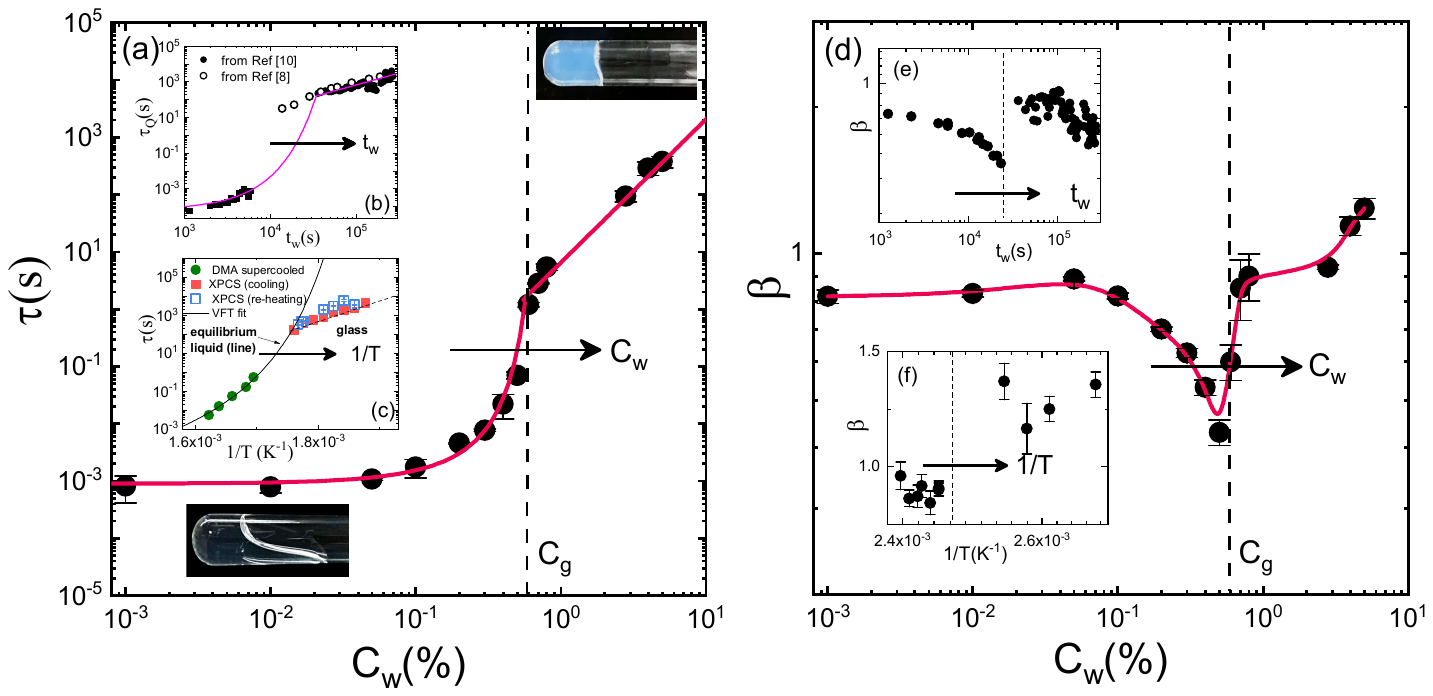}
\caption{(a) Structural relaxation time from Eq.(\ref{eq1}) as a function of weight concentration for IPN microgel suspensions at 
$C_{PAAc}$=24.6 \%, $T$=311 K and $Q$=0.022 nm$^{-1}$. Full lines
represent the best fits with Eq.(\ref{VFT}) and power law behaviours at small and large $C_w$ respectively. (b) Structural relaxation time behaviour with aging time $t_w$ from Ref.~\cite{AngeliniSM2013} for colloidal suspensions of Laponite\textsuperscript{\textregistered} at $C_w$=3.0 \% and (c) structural relaxation time behaviour with the inverse of temperature $1/T$ from Ref.~\cite{EvensonPRL2015} for a metallic glass. 
(d)  $\beta$ parameter from Eq.(\ref{eq1}) as a function of weight concentration for IPN microgel suspensions. Full line is a guide to eyes.
 (e) $\beta$ exponent as a function of aging time $t_w$ from Ref.~\cite{TudiscaRSC2012, AngeliniSM2013} for colloidal suspensions of Laponite\textsuperscript{\textregistered} at $C_w$=3.0 \% and (f) $\beta$ exponent dependence with $1/T$ from Ref.~\cite{RutaPRL2012} for a metallic glass.}
\label{tauBetaCw}
\end{figure*}

The corresponding structural relaxation time $\tau$ is reported in Fig.~\ref{tauBetaCw}(a). A similar behaviour of the viscosity has been observed for different types of soft colloids such as as block copolymers micelle, star polymers and microgels as described in Ref. ~\cite{VlassopoulosCOCIS2014} and references therein.
An interesting behaviour shows up in Fig. ~\ref{tauBetaCw}(a): at low
$C_w$, $\tau$ grows exponentially up to a critical weight concentration value $C_g$ with a behaviour $\tau=\tau_0 \: exp(\; \frac{D_{C_w} C_w}{C_{w0}-C_w})$. Above $C_g$ it suddenly increases as a power law  $\tau\sim
C_w^{\alpha}$ with $\alpha = 2.6 \pm 0.1$ signing the existence of two different dynamical regimes. A similar crossover has been recently reported for PNIPAM microgels ~\cite{PhilippePRE2018}  and share many similarities with the dynamics of hard colloids and structural glasses close to an out of equilibrium state.  
The relaxation time vs aging time $t_w$ of an hard colloid, the largely studied Laponite\textsuperscript{\textregistered} suspensions~\cite{BandyopadhyayPRL2004,TanakaPRE2005, RuzickaNatMat2011, AngeliniNC2014}, is reported in Fig.~\ref{tauBetaCw}(b) at $C_w$=3.0 \% from Ref.~\cite{AngeliniSM2013}. It evolves from an exponential behaviour $\tau=\tau_0 \: exp(\; \frac{D_t t_w}{t_w^{\infty}-t_w})$~\cite{BellourPRE2003, SchosselerPRE2006, AngeliniSM2013, AngeliniCSA2015} where $t_w$ drives the system out of equilibrium and $t_w^{\infty}$ sets the apparent divergence~\cite{RuzickaPRL2004, SahaSM2014}, to a slower than Arrhenius dependence with increasing $t_w$. Fig.~\ref{tauBetaCw}(c) shows instead the temperature dependence of the structural relaxation time for a metallic glass~\cite{EvensonPRL2015}. Also in this case, widely investigated in the last years~\cite{EvensonPRL2015, RutaPRL2012, RutaJPCM2017}, the existence of two dynamical regimes below and above the glass transition temperature $T_g$ has been observed. After $T_g$ the weaker temperature dependence of the structural relaxation time is well described by an
Arrhenius behaviour. It is worth to note that the dynamical crossover is reached by increasing concentration (Fig.~\ref{tauBetaCw}(a)) or waiting time (Fig.~\ref{tauBetaCw}(b)) in colloidal glasses and by decreasing temperature (Fig.~\ref{tauBetaCw}(c)) in structural glasses. 
Our findings for IPN microgels suggest that a unifying scenario can be provided: the slowing down of the dynamics achieved by varying weight concentration~\cite{LiSM2017,PhilippePRE2018, CipellettiPRL2000}, waiting time~\cite{BellourPRE2003, SchosselerPRE2006, AngeliniSM2013, AngeliniCSA2015} or temperature~\cite{RutaPRL2012, EvensonPRL2015} is accompanied by a super Arrhenius increase of the structural relaxation time followed by a slower than Arrhenius or an Arrhenius behaviour. Thus the existence of two different dynamical regimes is a general feature of soft colloids, hard colloids and structural glasses.

To complement these observations, we  report in Fig.~\ref{tauBetaCw}(d) the behaviour of the $\beta$ parameter derived from the fits through Eq.(\ref{eq1}). Surprisingly we find that the dynamical crossover at $C_g$ is signaled by a change of the shape of correlation functions from stretched to compressed with a corresponding sharp change of the shape parameter $\beta$ from $\beta<1$ for $C_w<C_g$ to $\beta>1$ for high concentrated samples that we attribute to rising stresses deriving from the high packaging of the particles.  As in the case of the structural relaxation time, the behaviour of $\beta$ displays many similarities with that observed in the hard colloid of Ref.~\cite{TudiscaRSC2012, AngeliniSM2013} as a function of $t_w$ (Fig.~\ref{tauBetaCw}(e)) and in the metallic glass of Ref.~\cite{EvensonPRL2015} as a function of 1/T (Fig.~\ref{tauBetaCw}(f)). For all these systems the dynamical crossover driven by $C_w$, $t_w$ or $T$ is always accompanied by a discontinuity of $\beta$, as also found for PNIPAM microgels ~\cite{PhilippePRE2018}, corroborating the existence of a universal behaviour regardless of the control parameter and the specific interactions of the system. It is worth to note that in the first dynamical regime $\beta$ decreases below 1 while in the second one it increases to values up to $\simeq$ 1 $\div$ 1.5 depending on the system. This deviation from the common stretched behaviour can be attributed to the development of internal stresses at the transition~\cite{CipellettiPRL2000, RutaPRL2012, AngeliniSM2013, BouzidNatCom2017, WeiNatCom2018}. The peculiar values of each system depend on the specific material, on its history and/or on the preparation protocols whose optimization and control is subject of theoretical and experimental investigations.  

\begin{figure}
\centering
\includegraphics[width=8cm]{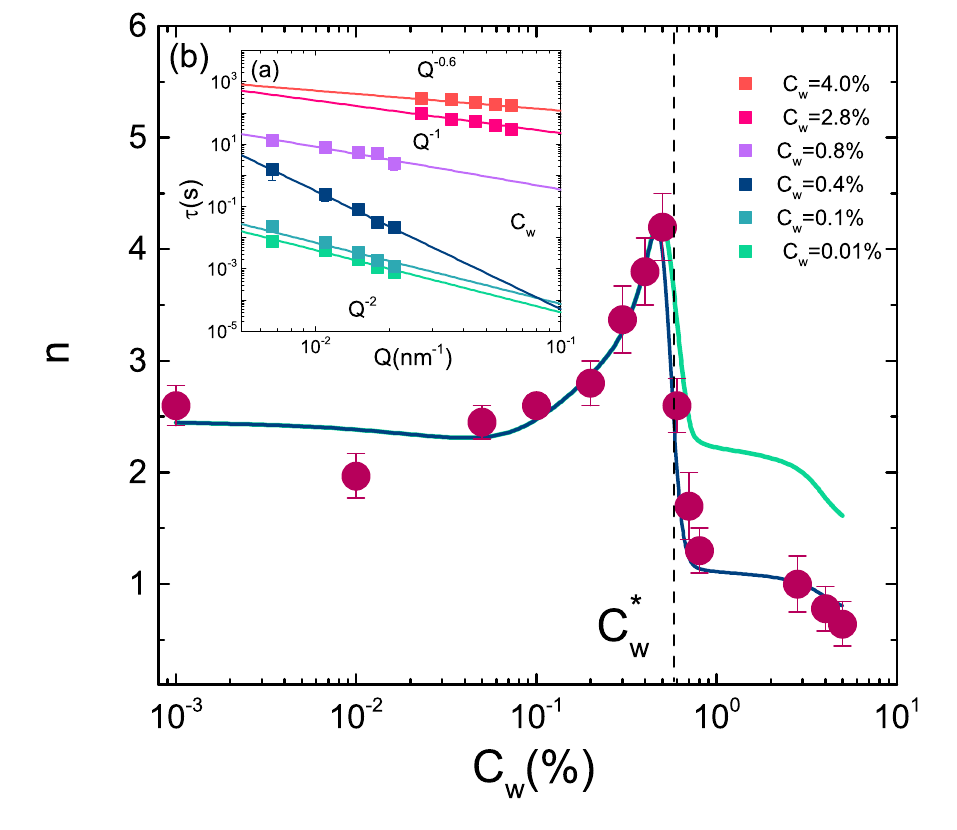}
\caption{(a) Structural relaxation time as a function of the scattering vectors $Q$ for IPN microgel suspensions  
at $C_{PAAc}$=24.6 \% and $T$=311 K as measured through DLS and XPCS at increasing weight concentrations. Full lines represent the fits through Eq.\ref{powerlaw}. (b) Power exponent from fits trough Eq.\ref{powerlaw}. Blue full line is the predicted behaviour from Eq.\ref{Eq:Gauss} while green full line is the predicted behaviour using Eq.\ref{tauCw1}.}
\label{tauQ}
\end{figure}

The $Q$-dependence of the structural relaxation time in the range $Q$=(0.006 $\div$ 0.063) nm$^{-1}$ is reported in Fig.~\ref{tauQ}(a) at different concentrations. The data are well fitted through a power law 

\begin{equation}
\tau(Q) \propto Q^{-n}  \label{powerlaw}
\end{equation}

shown as solid lines in Fig.~\ref{tauQ}(a). The behaviour of $n$  as a function of weight concentration is reported in Fig.~\ref{tauQ}(b). Two dynamical regimes can be distinguished below and above $C_g$. In the first one ($C_w$ < $C_g$) $n$ increases from values typical of diffusive dynamics ($n \sim 2$), to unusual values ($n \sim 4$) for which diffusive dynamics is ruled out.

In the second dynamical regime ($C_w$ $\geq$ $C_g$)  $n$ suddenly decreases down to values $n \leq	1$ where the case of $n \sim$1 is peculiar of the ballistic motion of particles. The crossover between the two dynamical regimes is characterized by a maximum at the weight concentration $C_g$ and is also associated to the observed crossover of the structural relaxation time (Fig.~\ref{tauBetaCw}(a)) and $\beta$ parameter (Fig.~\ref{tauBetaCw}(d)).
Aiming to explain this behaviour we consider that at low weight concentration $C_w<0.1$ \% particles freely  move, the diffusion process is well described by Fickian law, as in the case of Brownian motion, and the mean square displacement $<r^2(t)>$ has a linear time dependence:

\begin{equation} 
<r^2(t)>= 6D t \label{MSD1}
\end{equation}

where $D$ is the particle diffusion coefficient.  
At increasing weight concentraton $C_w>0.1$ \% the system starts to become spatially heterogeneous, particles are no longer free to move as in the case of pure diffusive dynamics since they are trapped in cages characterized by different persisting times. In our experiments we probe scattering vectors $Q$ in the range $Q$=(0.006 $\div$ 0.063) nm$^{-1}$ that corresponds to length scales up to ten particles radii (R=(130 $\pm$ 2) nm) and we are thus sensitive to these heterogeneities. As a consequence Eq.~\ref{MSD1} fails to reproduce the data which instead can be described by a non-Fickian anomalous diffusion~\cite{HavlinAdvPhys1987}: 

\begin{equation}
<r^2(t)> = \Gamma t^{\beta} \label{MSD2}
\end{equation}

with an effective diffusion coefficient~\cite{SokolovSM2012, PalomboSciRep2013} $D(t)=\Gamma t^{\beta -1 }$. The \textit{diffusive dynamics} (Eq.~\ref{MSD1}) is recovered when \textit{$\beta=1$} and $D(t)=\Gamma=6D$.
For \textit{$0<\beta<1$} the dynamics is \textit{subdiffusive} and for \textit{$\beta> 1$} is \textit{superdiffusive}, with the special case of \textit{$\beta=2$} corresponding to \textit{ballistic} motion.

The mean square displacement, in the Gaussian approximation, is related to the intermediate scattering function as \cite{HansenMcDonald2013}:  

\begin{equation} 
F(Q,t)=exp(-Q^2<r^2(t)>/6) \label{Gaussian}
\end{equation}

that, using the generalized mean square displacement of Eq.~\ref{MSD2}, becomes $F(Q,t)=exp{(- \frac{Q^2 \Gamma}{6} t^{\beta})}$. The comparison with the KWW expression (Eq.(\ref{eq1})) of the intermediate scattering function $F(Q,t) \propto exp{(-(t/\tau(Q))^{\beta})}$, gives $(t/\tau(Q))^{\beta}=\frac{Q^2 \Gamma}{6} t^{\beta}$ and:

\begin{equation} 
(1/\tau(Q))^{\beta}=Q^2 \Gamma/6 
\end{equation}
that provides:

\begin{equation} 
\tau(Q) \propto Q^{-2/\beta} \label{tauCw1}
\end{equation}
 
In the case of diffusive dynamics with $\beta$=1 one gets the typical $\tau(Q) \propto Q^{-2}$ behaviour while in the case of ballistic motion of particles with $\beta$=2 $\tau(Q) \propto Q^{-1}$. A comparison between Eq.\ref{powerlaw} and Eq.\ref{tauCw1} gives $n=2/\beta$. This behaviour is reported in Fig.\ref{tauQ}(b) as green line. One can observe a good agreement with the experimental data only for $C_w<C_g$. The weight concentration $C_g$, characterized by a relaxation time of the order of seconds (Fig.\ref{tauBetaCw}(a)), sets the transition from a fluid to an intermediate state that is visually arrested, as shown in the top photograph of Fig.\ref{tauBetaCw}. For $C_w \geq C_g$ the dynamics slows down progressively towards another arrested state, as witnessed by the increasing relaxation times up to values of the order of thousand seconds. Here the Gaussian approximation of Eq.\ref{Gaussian} becomes inaccurate~\cite{BalucaniZoppi1995} and a different scenario has to be invoked. The dynamics approaching this state can be described whitin the Mode Coupling Theory (MCT)~\cite{BennemannEPJB199, Gotze2009} that provides a different $Q$-dependence of the structural relaxation time:

\begin{equation}
\tau(Q) \propto Q^{-1/\beta} \label{tauCw2}
\end{equation}

Surprisingly for $C_w \geq C_g$ a perfect agreement is now observed as shown in Fig.\ref{tauQ}(b) (blue full line).
Therefore the experimental data are well described through:

\begin{eqnarray} 
\tau(Q) \propto Q^{-2/\beta}  \qquad C_w<C_g \nonumber  \\
\tau(Q) \propto Q^{-1/\beta}   \qquad C_w\geq C_g 
\label{Eq:Gauss}
\end{eqnarray}

\section{Conclusions}

In conclusion through XPCS and DLS measurements on IPN microgels we show that interpenetrating different amount of PAAc into PNIPAM network originates particles with varying softness allowing to finely control the fragility of the system. Moreover, the fundamental questions opened in the introduction have been addressed. In particular we find the existence of two dynamical regimes, with a crossover of the \textit{structural relaxation} time associated to a clear change of the 
shape of correlation functions from stretched to compressed with a corresponding sharp change of the shape parameter $\beta$ from $\beta<1$ to $\beta>1$ . The comparison with similar behaviours found in hard colloids and in metallic glasses indicates that these are \textit{general common features of different systems} regardless of the \textit{control parameters} ($C_w$, $t_w$, T) and of the specific interactions at stakes.
The scattering vector dependence of the structural relaxation time in IPN microgels displays a surprising behaviour of the derived power exponent with a maximum in correspondence of the dynamical crossover. We provide a phenomenological description of the experimental data: at very low concentration, when particles are free to move, \textit{the dynamics} is well described by a Fickian diffusion, at intermediate concentrations, when particles start to be trapped in cages, an effective non Fickian anomalous diffusion 
has to be considered, while at the highest investigated concentrations, when the systems is going towards another arrested state, the emergence of a ballistic motion is well described within the Mode Coupling Theory.
All these results contribute to the further understanding of the role of softness in the fragility, to the analogies and differences between colloidal and structural glasses and add new insights to the so far debated stretched to compressed transition. In order to further test our findings
and hypothesis, more experiments on a wide range of systems will be useful.

\section*{Acknowledgments}
The authors acknowledge acknowledge ESRF for beamtime and support from MIUR Fare SOFTART (R16XLE2X3L).

\subsection*{Author Contributions}
R.A., V.N. and B.Ruz. conceived the experiments. R.A., V.N., B.R., B.Ruz. and F.Z. conducted the experiments. R.A., V.N. and B.Ruz. analysed the results. M.B. and E.B. synthesized the samples. R.A., V.N. and B.Ruz. wrote the manuscript.


%

\end{document}